# Electronic Transitions in Strained SmNiO$_3$ Thin Films

S. Catalano,[1,a)] M. Gibert,[1] V. Bisogni,[2,b)] O. Peil,[1] F. He,[3] R. Sutarto,[3] M. Viret,[1,4] P. Zubko,[1,c)] R. Scherwitzl,[1] A. Georges,[1,5] G. A. Sawatzky,[6] T. Schmitt,[2] J.-M. Triscone[1]

[1]*Department de Physique de la Matière Condensée, Université de Genève, Genève, 1211, Switzerland*
[2]*Swiss Light Source, Paul Scherrer Institut, CH-5232, Villigen, Switzerland*
[3]*Canadian Light Source, Saskatoon, Saskatchewan, S7N 2V3, Canada*
[4]*Service de Physique de l'Etat Condensé (CNRS URA 2464), CEA Saclay, 91191, Gif-sur-Yvette, France*
[5]*Centre de Physique Théorique (CPHT), École Polytechnique, Palaiseau, Cedex, 91128, France*
[6]*Department of Physics and Astronomy, University of British Columbia, Vancouver, British Columbia, V6T 1Z1, Canada*

Nickelates are known for their metal to insulator transition (MIT) and an unusual magnetic ordering, occurring at $T=T_{Néel}$. Here, we investigate thin films of SmNiO$_3$ subjected to different levels of epitaxial strain. We find that the original bulk behavior ($T_{Néel}<T_{MI}$) is strongly affected by applying compressive strain to the films. For small compressive strains, a regime where $T_{Néel}=T_{MI}$ is achieved, the paramagnetic insulating phase characteristic of the bulk compound is suppressed and the MIT becomes 1$^{st}$ order. Further increasing the in-plane compression of the SmNiO$_3$ lattice leads to the stabilization of a single metallic paramagnetic phase.

[a)] Author to whom correspondence should be addressed. Electronic mail: sara.catalano@unige.ch
[b)] Currently at National Synchrotron Light Source II, Brookhaven National Laboratory, Upton, New York, 11973, USA.
[c)] Currently at London Centre for Nanotechnology and Department of Physics and Astronomy, University College London, 17-19 Gordon Street, London WC1H 0HA, UK.

Transition metal oxides (TMO) constitute the essential building blocks of novel functional devices due to their unique physical properties, which arise from the interplay of the TM orbitals and the surrounding oxygen network.[1]

In the case of perovskite nickelates (RE-NiO$_3$, RE=Rare Earth), except for RE=La, a metal to insulator transition (MIT) is observed as the temperature is reduced below $T=T_{MI}$.[2,3] The MIT is controlled by the distortion of the unit cell (u.c.) with respect to an ideal cubic structure. The Ni-O-Ni angle ($\Phi$) deviates from 180° and therefore tunes the hopping integral between the O $2p$ and Ni $3d$ orbitals, which in turn governs the effective bandwidth ($W$) of the system.[4] $\Phi$ depends on the $R$ radius or, equivalently, on the perovskite tolerance factor $t$.[4,5] Exploiting a variety of tools, such as chemical substitution,[2,3,4] hydrostatic pressure[5-7] and, in the case of thin films, epitaxial strain,[8-10] it has been shown that $T_{MI}$ monotonically decreases as $\Phi$ approaches 180°, and eventually only the metallic phase is observed for sufficiently small structural distortion.[5,7,10-12]

Concomitant to the MIT, a change of symmetry from the orthorhombic $Pbnm$ to the monoclinic $P2_1/n$ space group is reported.[13] This symmetry lowering has been associated with the occurrence of a breathing distortion of the oxygen octahedra, establishing two inequivalent nickel sites.[13,14,15,16] Finally, RE-NiO$_3$ (RNOs) undergo a paramagnetic (PM) to antiferromagnetic (AFM) transition at $T=T_{Néel}$. The AFM structure gives rise to a magnetic diffraction peak with the Bragg vector $q_{Bragg}=(¼ ¼ ¼)$, in *pseudo-cubic* (*pc*) notation.[2,3,17,18] The magnetic phase seems to arise from the bond disproportionated insulating phase. Depending on $R$, the temperature of the Néel transition reveals two distinct regimes in the nickelate phase diagram. The members with $R$ smaller than Nd display a 2$^{nd}$ order magnetic transition at $T_{Néel}<T_{MI}$, where the Néel temperature increases slightly with $R$.[18] In contrast, for Nd and larger $R$, $T_{Néel}=T_{MI}$ and the magnetic ordering shares the $R$ dependence of the MIT and its 1$^{st}$ order character.[2,17]

Most of the experimental work performed on thin films has been focused on the properties of RNO with $T_{Néel}=T_{MI}$, partly due to the difficulties in synthesizing compounds with a more distorted ($R<$Nd) structure.[19] However, interest in more distorted RNO has been renewed by recent results,[19-22] confirming their potential for both fundamental physics[21] and applied research.[22] Here, we investigate the properties of thin films of SmNiO$_3$ (SNO), which in its bulk form has $T_{MI}=400$ K and $T_{Néel}=225$ K,[3] and is representative of the RNO with $T_{Néel}<T_{MI}$. By selecting a variety of single crystal substrates, we impose different levels of epitaxial strain ($\varepsilon_{xx}$) to the films. We use transport measurements and Resonant Soft X-Ray Diffraction (XRD) to determine the onset of the MI and Néel transitions in the samples, as a function of the applied in-plane strain. A crossover from the $T_{Néel}<T_{MI}$ to $T_{Néel}=T_{MI}$ regime is observed as compressive strain is increased, eventually stabilizing a unique PM metallic phase. These findings are compared to the case of epitaxially strained thin films of NdNiO$_3$ (NNO), which is representative of the RNO with $T_{Néel}=T_{MI}$, and summarized on a strain-temperature phase diagram.

Coherently strained SNO epitaxial thin films were grown by off-axis radio frequency magnetron sputtering on a variety of [001]$_{pc}$-oriented single crystal substrates, providing strain ranging from $\varepsilon_{xx}=+1.9\%$ to $\varepsilon_{xx}=-2.3\%$. Optimal deposition conditions were found to be at a total pressure of 0.18 mbar with an oxygen/argon mixture of 1:3, and a substrate temperature of 460° C. Prior to deposition, all substrates were thermally treated to ensure atomically flat terraces and step-like morphology. The thickness of the films was about 25-30u.c. in order to preserve the strain across the entire layer. Details of the structural parameters of the films are reported in Ref. 23.

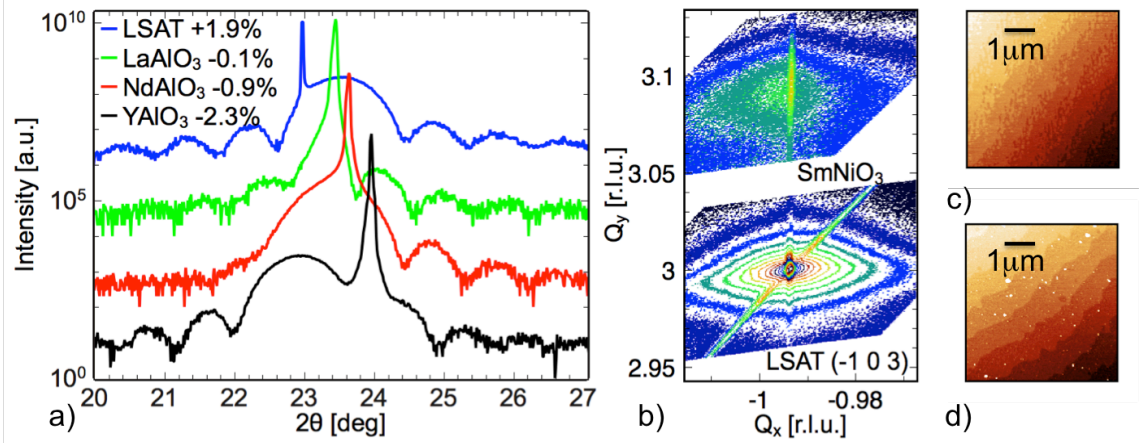

**FIG. 1** Structural characterization of 30 u.c. strained SNO films. (a) Symmetric θ/2θ scans for films grown under different strains, demonstrating high crystallinity of the samples. (b) Reciprocal space map around the $(-103)_{pc}$ reflection for a film grown on LSAT: the intensity of the film reflection is mainly concentrated at the $Q_x$ coordinate of the substrate, showing that the films are fully strained to the substrate lattice even for high strain level ($\varepsilon_{xx}$ = +1.9%). The axes correspond, respectively, to the $h$ ($Q_x$) and $l$ ($Q_y$) directions of reciprocal space. (c),(d) Typical atomic force microscopy images of SNO grown on top of (c) $YAlO_3$, and (d) $LaAlO_3$.

Figure 1 presents a typical *ex situ* characterization of the SNO films, exemplifying the quality of the samples. X-ray diffraction scans, reciprocal space mapping (Q-map) and topography images attest to the monocrytsalline and coherently strained structure of the films and their atomically flat surfaces. High quality NNO films were deposited and characterized with analogous methods, as reported in Refs. 12 and 23.

The resistivity (ρ) of the strained SNO films is plotted as a function of temperature in Figure 2. The film deposited on top of $LaAlO_3$, which provides a negligible compressive strain ($\varepsilon_{xx}$=-0.1%), displays a MIT at 380 K. As in the bulk, no hysteresis is observed as the temperature is cycled back and forth between 4 K and 400 K. However, applying larger compressive strains dramatically affects the MIT. For $\varepsilon_{xx}$=-0.9 % (SNO/$NdAlO_3$), $T_{MI}$ drops well below room temperature ($T_{MI}$≈144 K) and ρ exhibits a hysteresis loop typical of a 1$^{st}$ order transition, with resistivity changes of over two orders of magnitude. For $\varepsilon_{xx}$=-2.3% (SNO/$YAlO_3$), the insulating phase is completely suppressed and the film shows metallic behavior down to the lowest measured temperature (*T*=4 K).

The gradual decrease of $T_{MI}$ for increasing negative strain indicates that the in-plane compression results in a broadening of the bandwidth of the system,[24] suggesting that the SNO lattice accommodates the epitaxial constraint by reducing the Ni-O-Ni bending. Indeed, DFT calculations performed in order to determine the crystal structure of coherently strained SNO reveal that imposing a reduced in-plane lattice constant induces a straightening of the out-of-plane Ni-O-Ni angle ($\Phi_z$) and leaves the in-plane Ni-O-Ni angle ($\Phi_{xy}$) almost unaffected.[23] Such behavior is reminiscent of the response of epitaxial NNO films to compressive strain[8-10, 23] and of the bulk compounds to hydrostatic pressure,[5-7] which also exhibit a gradual reduction of $T_{MI}$ as a consequence of a decreased degree of structural distortion. Thus, epitaxial strain is an efficient tool for tuning the RNO bandwidth, also for a member with *R*<Nd. Remarkably, the observed shift of the MIT is much larger than has been achieved by applying hydrostatic pressure on bulk SNO,[6] highlighting the advantage of synthesizing RNO films for tuning of their functional properties. The crossover from a 2$^{nd}$ order to a 1$^{st}$ order MIT is even more striking, as no analogous change of the MIT character has been reported by applying epitaxial strain to NNO. Here, for a relatively small amount of compression, SNO films suddenly display the transport properties of RNOs with $R \geq Nd$. It is interesting to note that the room temperature resistivity of the most compressively strained SNO films (SNO/YAlO$_3$) is unexpectedly larger than the resistivity of the metallic phase of the films subjected to smaller strain levels (Fig. 2). This result, that may apparently contradict the increase of the bandwidth predicted through DFT calculations, reveal that other more subtle effects should be taken into account. For example, a significant role can be played by the symmetry of the substrate, as the discrepancy is observed between the resistivity of the metallic phase of SNO films grown on LaAlO$_3$, NdAlO$_3$ (both rhombohedral) and films grown on YlAlO$_3$ (orthorhombic). In addition, we suggest that other experimental factors, such as the quality of the substrates, could lead to different scattering rates at the film/substrate interface, therefore affecting the resistivity of the films. Finally, it is important to mention that LaNiO$_3$ (only RNO with $T_{MI}=0$ in bulk) is also the only member of the RNO family having a rhombohedral structure (space group $R3c$). Therefore, it would be interesting to study the structure of strained SNO and NNO films with $T_{MI}=0$, and verify whether their structure is rhombohedral too.

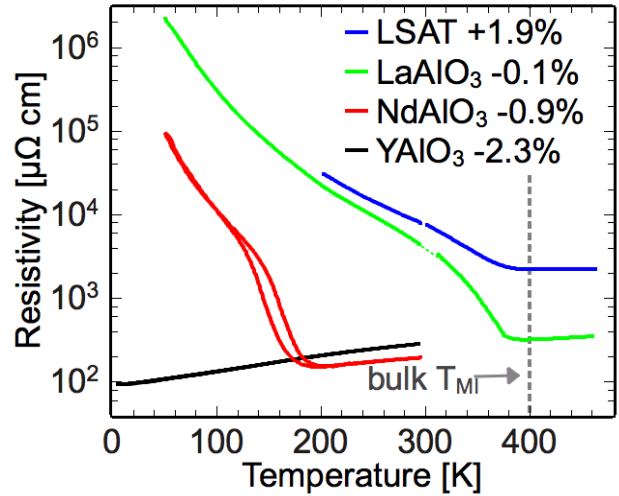

**FIG. 2** Temperature dependence of resistivity for SNO films subjected to different degrees of strain. The dashed line indicates the $T_{MI}$ reported for bulk SNO.[3]

By depositing SNO on LSAT ($\varepsilon_{xx}=+1.9\%$, blue curve in Figure 2) we also investigate the impact of strain in the tensile regime. The MIT occurs at $T_{MI}=390K$ and is non-hysteretic, just as for SNO on LaAlO$_3$. However, compared to other compressively strained SNO films, the transition is broader and the resistivity of the metallic phase is almost an order of magnitude higher. These observations can be ascribed to the creation of oxygen vacancies in the films as a mechanism to release the tensile strain, which has been extensively discussed in Ref. 25 and Ref. 26. This interpretation is supported by the anomalously large increase of the unit cell volume observed for the SNO film under tensile strain.[23]

In order to inspect the magnetic properties of the films, we performed resonant soft x-ray diffraction measurements at the RSXS endstation of the REIXS beamline at the Canadian Light Source (CLS).[27] For a description of the experimental set-up see Ref. 23. We probed the presence of the (¼ ¼ ¼) Bragg reflection, corresponding to the AFM phase of the bulk compounds, in strained SNO and NNO films as the temperature was varied from 20K to 300K. Figure 3(a) illustrates the strong sharp peak detected in SNO/LaAlO$_3$ at $T$=20K, corresponding to the mentioned q$_{Bragg}$, pointing to a 4 u.c periodic superstructure oriented along the [111]-axis in the low temperature phase of the sample. The diffraction peak resonates when the incoming beam energy is tuned to the Ni $L_{2,3}$ and RE $M_{4,5}$ edges, confirming that the signal originates from the RNO lattice. By changing the polarization of the incident beam, a strong dichroism between the σ and π component (black and red line in Figure 3(a)) is observed, as expected for diffraction of magnetic origin, where the σ-σ scattering channel is prohibited.[28] Taking into account the resonant nature of the peak, the observed dichroic effect and the clear similarity to the bulk case, we can reasonably state that the detected Bragg reflection is of magnetic origin and is due to the AFM ground state of the SNO film. An analogous resonant peak was present in the low temperature insulating phase of SNO/NdAlO$_3$[23] and NNO/LaAlO$_3$ (Figure 3(c)). In contrast, no (¼ ¼ ¼) Bragg reflection was detected in the films showing only a metallic phase (SNO/YAlO$_3$; NNO/NdAlO$_3$), indicating that they remain paramagnetic over the considered temperature range.

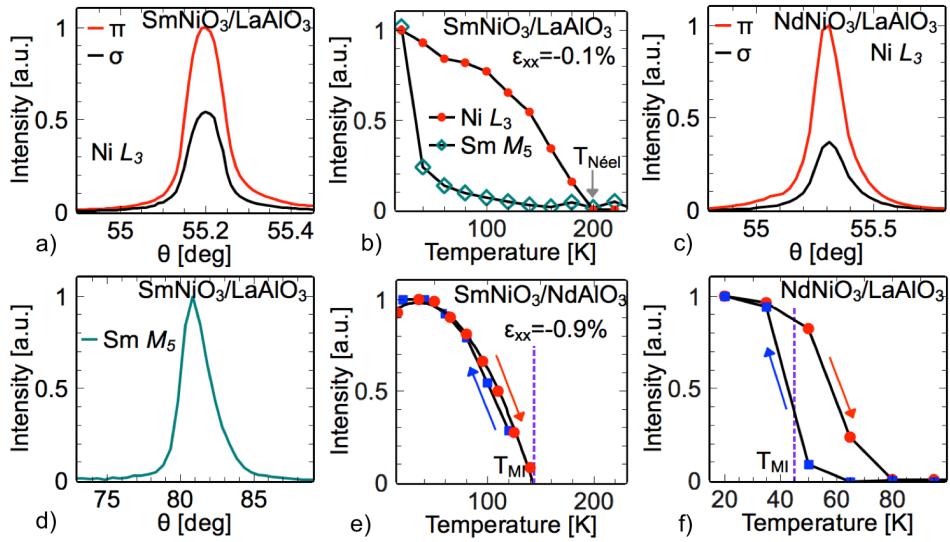

**FIG. 3** Magnetic properties of the films characterized by resonant X-ray diffraction. (a) Rocking curve of the SNO/LaAlO$_3$ film about q$_{Bragg}$=(¼ ¼ ¼), at T=20K, measured with π (red line) and σ (black line) linearly polarized light. (b) Temperature dependence of the peak intensity measured at the Ni $L_3$ (red circles) and at the Sm $M_5$ (pink diamonds) edge. (c) Rocking curve of NNO/LaAlO$_3$ film around q$_{Bragg}$=(¼ ¼ ¼), at T=20K. (d) Rocking curve about q$_{Bragg}$=(¼ ¼ ¼) measured at the Sm $M_5$ resonance. Bragg peak intensity of (e) SNO/NdAlO$_3$, and (f) NNO/LaAlO$_3$, measured at the Ni $L_3$ edge, as the temperature was cycled up (red circles) and down (blue squares). The vertical dashed lines indicate the T$_{MI}$ of the corresponding films.

Figures 3(b),(e) show the evolution of the magnetic diffraction peak intensity as a function of temperature in the SNO films, as the photon energy was tuned to the Ni $L_3$ edge. The intensity *versus* temperature curves were obtained by integrating rocking curve scans acquired at different temperatures, the blue (red) symbols corresponding to the cooling (heating) leg of the cycle. We define T$_{Néel}$ of the films as the temperature at which the diffraction intensity vanishes, determined by extrapolation. The temperature dependence of the magnetic peak intensity of SNO/LaAlO$_3$ (Figure 3(b)) displays the typical behavior of the order parameter of a system with a 2$^{nd}$ order transition, developing at T$_{Néel}$ ≈ 200 K, fairly close to the value reported for bulk SNO.[3] In the case of SNO/NdAlO$_3$ (Figure3(e)), the intensity increases abruptly at T$_{Néel}$≈140 K, matching the above reported T$_{MI}$ of the film. The measurements reveal that hysteresis is present between the heating and cooling parts of the cycle, reproducing the corresponding ρ *versus* temperature curve (Figure 2) and pointing to a 1$^{st}$ order character of the Néel transition. The 1$^{st}$ order character of the AFM ordering was observed in the NNO films as well (Figure 3(f)), confirming previously reported results.[2,4,29,30] Figures 3(b),(d) present the temperature dependence of the (¼ ¼ ¼) reflection detected at the Sm $M_5$ edge of SNO/LaAlO$_3$. This measurement reveals that the ordering of the RE magnetic sublattice follows the polarization of the Ni moments (Figure 3(b)), bearing out its induced nature, as already observed in the bulk nickelate[4] and in NNO thin films.[29] The same behavior at the RE resonance was reproduced in all the SNO and NNO films investigated.[23]

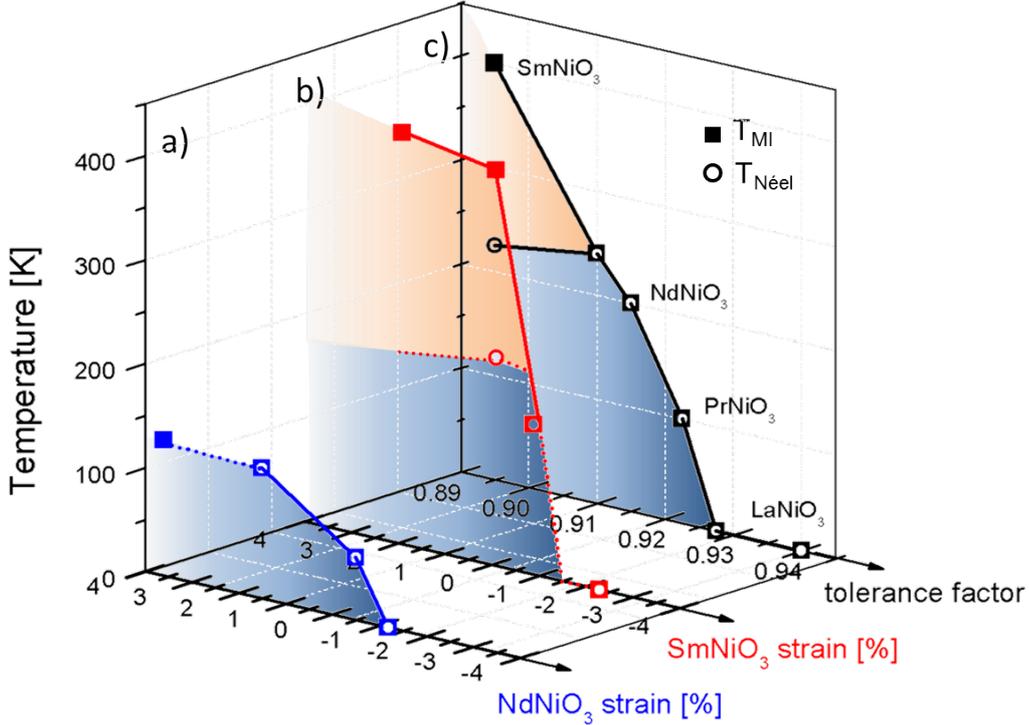

**FIG. 4** Phase diagram of nickelates summarizing the evolution of $T_{MI}$ and $T_{Néel}$ of strained (a) NNO and (b) SNO films, compared to the bulk compounds (c). Solid lines are guides to the eye, and dashed lines represent the expected behavior for $T_{Néel}$ of SNO/LSAT ($\varepsilon_{xx}$= +1.9%) and NNO/DyScO$_3$ ($\varepsilon_{xx}$=+3.6%). The x-axis of panel (a) and (b) goes from positive to negative values in order to facilitate the comparison with panel (c). The data for the bulk RNOs have been extracted from Ref. 3 and Ref. 12.

From our observations, we can draw the following conclusions. First, the AFM phase only develops if an insulating phase of the RNO layer exists, that is, if electronic localization occurs. Compressive strain suppresses the PM insulating phase of SNO, suggesting that the Néel and MI transitions couple once the bandwidth is sufficiently broad to drive $T_{MI}$ below the intrinsic $T_{Néel}$ of the compound. Second, the 1$^{st}$ order behavior of the MIT occurs when $T_{MI}$=$T_{Néel}$, showing that the dynamic of the transitions is affected by the interplay of magnetic and charge fluctuations. Furthermore, the absence of any signature of magnetic ordering in the metallic phase of the samples, down to the lowest *T* measured (*T*=4 K), demonstrates that the magnetic coupling is effectively screened by the charges in this itinerant regime resulting in too weak magnetic interactions to allow for long range spin ordering.

Figure 4 summarizes the evolution of the MI and Néel transitions of NNO and SNO films (first and second panel) as a function of strain, compared to the dependence on tolerance factor *t* observed for the bulk compounds. The phase diagram clearly illustrates that epitaxial strain allows tuning the electronic properties of SNO films from the regime with $T_{Néel}<T_{MI}$ to the case of $T_{Néel}=T_{MI}$, including the extreme case of $T_{MI}=T_{Néel}=0$ K. Thus the phase diagram of epitaxially strained SNO films is qualitatively the same as that of the bulk nickelates with different compositions, with biaxial strain playing the same role as the tolerance factor $t$. By contrast, the NNO films show a single electronic regime with $T_{Néel}=T_{MI}$.

In conclusion, we have investigated the evolution of the electronic phases of strained nickelate thin films as a function of temperature and shown that, by exploiting epitaxial strain, we can control the MI and Néel transitions of SNO films. Our theoretical calculations show that compressive strain is accommodated by straightening the Ni-O-Ni angle, whereas tensile strain has been associated with an increased content of oxygen defects. By changing $\Phi$, the bandwidth of the system is modified, leading to a continuous shift of the MIT from $T_{MI}=400$ K down to $T_{MI}=0$ K. Noticeably, the MIT of SNO films can be reduced by over 200 K by applying just $\varepsilon_{xx}=-0.9\%$. From resonant X-ray diffraction measurements, we have direct evidence of the as yet unreported AFM ordering in epitaxially strained SNO films, with an AFM structure akin to that of the bulk. The measurements reveal that the AFM transition only takes place in the insulating phase of the system, i.e. the detected $T_{Néel}$ is never above $T_{MI}$, as already observed for the bulk compounds. The use of resonant excitation with soft X-rays also allows us to distinguish the role played by the Ni and $R$ magnetic moments in the evolution of the AFM phase, showing that the ordering is triggered by the polarization of the Ni spins.

**ACKNOWLEDGMENTS**

We thank Marco Lopes and Sebastian Muller for their precious support to the experiments. We also acknowledge Siobhan McKeown Walker and Jennifer Fowlie for proofreading the manuscript. This work was supported by the Swiss National Science Foundation through the National Center of Competence in Research, Materials with Novel Electronic Properties, 'MaNEP' and division II. The research leading to these results has received funding from the European Research Council under the European Union's Seventh Framework Programme (FP7/2007-2013) / ERC Grant Agreement n° 319286 and under grant agreement Nr. 290605 (COFUND: PSI-FELLOW). Part of the research described in this paper was performed at the Canadian Light Source, which is funded by the CFI, NSERC, NRC, CIHR, the Government of Saskatchewan, WD Canada and the University of Saskatchewan.

# REFERENCES

[1] E. Dagotto, *Science* **309**, 257 (2005); J. W. Reiner, F. J. Walker, C. H. Ahn, *Science* **323**, 1018 (2009); H. Takagi, H. I. Hwang, *Science* **327**, 1601 (2010).

[2] P. Lacorre, J. B. Torrance, J. Pannetier, A.I. Nazzal, P.W. Wang, T.C. Huang, *J. Solid State Chem.* **91**, 225 (1991).

[3] J. B. Torrance, P. Lacorre, A.I. Nazzal, E. J. Ansaldo, Ch. Niedermayer, *Phys. Rev. B* **45**, 8209(R) (1992).

[4] García-Muñoz, J. Rodríguez-Carvajal, P. Lacorre, J. B. Torrance, *Phys. Rev. B* **46**, 4414 (1992).

[5] P. C. Canfield, J. D. Thompson, S-W. Cheong, L. W. Rupp, *Phys. Rev. B* **47**, 357(R) (1993).

[6] J.-S. Zhou, J. B. Goodenough, B. Dabrowski, *Phys. Rev. Lett.* **95**, 127204 (2005).

[7] X. Obradors, L. M. Paulius, M. B. Maple, J. B. Torrance, A. I. Nazzal, J. Fontcuberta, X. Granados, *Phys. Rev. B* **47**, 12353 (1993).

[8] R. Scherwitzl, P. Zubko, I. G. Lezama, S. Ono, A. F. Morpurgo, G. Catalan, J.-M. Triscone, *Adv. Mater.* **22**, 5517 (2010).

[9] J. Liu, M. Kareev, B. Gray, J. W. Kim, P. Ryan, B. Dabrowski, J. W. Freeland, J. Chakhalian, *Appl. Phys. Lett.* **96**, 233110 (2010); Y. Kumar, R. J. Choudhary, S. K. Sharma, M. Knobel, Ravi Kumar *App. Phys. Lett.* **101**, 132101 (2012); J. Liu, M. Kargarian, M. Kareev, B. Gray, P. J. Ryan, A. Cruz, N. Tahir, Y.-D. Chuang, J. Guo, J. M. Rondinelli, J. W. Freeland, G. A. Fiete, J. Chakhalian, *Nat. Commun.* **4**, 2714 (2013); A. S. Disa, D. P. Kumah, J. H. Ngai, E. D. Specht, D. A. Arena, F. J. Walker, C. H. Ahn, *APL Mater.* **1**, 032110 (2013); D. P. Kumah, A. S. Disa, J. H. Ngai, H. Chen, A. Malashevich, J. W. Reiner, S. Ismail-Beigi, F. J. Walker, C. H. Ahn, *Adv. Mater* **26**, 1935 (2014).

[10] P.-H. Xiang, N. Zhong, C.-G. Duan, X. D. Tang, Z. G. Hu, P. X. Yang, Z. Q. Zhu, and J. H. Chu, *J. Appl. Phys.* **114**, 243713 (2013).

[11] M. L. Medarde, *J. Phys.: Condens. Matter* **9**, 1679 (1997).

[12] G. Catalan, *Phase Transitions* **81**, 729 (2008).

[13] J. A. Alonso, J. L. García-Muñoz, M. T. Fernández-Díaz, M. A. G. Aranda, M. J. Martínez-Lope, M. T. Casais, *Phys. Rev. Lett.* **82**, 3871 (1999); J. A. Alonso, M. J. Martinez-Lope, M. T. Casais, J. L. Garcia-Munoz, M. T. Fernández-Díaz, *Phys. Rev. B* **61**, 1756 (2000); M. Zaghrioui, A. Bulou, P. Lacorre, P. Laffez, *Phys. Rev. B* **64**, 081102(R) (2001); e) U. Staub, G. I. Meijer, F. Fauth, R. Allenspach, J. G. Bednorz, J. Karpinski, S. M. Kazakov, L. Paolasini, F. d'Acapito, *Phys. Rev. Lett.* **88**, 126402 (2002); M. Medarde, C. Dallera, M.

Grioni, B. Delley, F. Vernay, J. Mesot, M. Sikora, J. A. Alonso, M. J. Martínez-Lope, *Phys. Rev. B* **80**, 245105 (2009); S. Johnston, A. Mukherjee, I. Elfimov, M. Berciu, G. A. Sawatzky, *Phys. Rev. Lett.* **112**, 106404 (2014).

[14]I. I. Mazin, D. I. Khomskii, R. Lengsdorf, J. A. Alonso, W. G. Marshall, R. M. Ibberson, A. Podlesnyak, M. J. Martinez-Lope, M. M. Abd-Elmeguid, *Phys. Rev. Lett.* **98**, 176406 (2007).

[15]H. Park, A. J. Millis, and C. A. Marianetti, *Phys. Rev. Lett.* **109**, 156402 (2012).

[16]T. Mizokawa, D. I. Khomskii, G. A. Sawatzky, *Phys. Rev. B* **61**, 263 (2000)

[17]J. L. García-Muñoz, J. Rodríguez-Carvajal, P. Lacorre, *Phys. Rev. B* **50**, 978 (1994).

[18]J. Rodríguez-Carvajal, S. Rosenkranz, M. Medarde, P. Lacorre, M. T. Fernández-Díaz, F. Fauth, V. Trounov, *Phys. Rev. B* **57**, 456 (1998).

[17]I. Vobornik, L. Perfetti, M. Zacchigna, M. Grioni, G. Margaritondo, J. Mesot, M. Medarde, P. Lacorre, *Phys. Rev. B* **60**, 8426(R) (1999).

[18]C. Girardot, J. Kreisel, S. Pignard, N. Caillault, and F. Weiss, *Phys. Rev. B* **78**, 104101 (2008).

[19]J. A. Alonso, M. J. Martínez-Lope, M. T. Casais, M. A. G. Aranda, M. T. Fernández-Díaz, *J. Am. Chem. Soc*. **121**, 4754 (1999).

[20]D. Meyers, E. J. Moon, M. Kareev, I. C. Tung, B. A. Gray, J. Liu, M. J. Bedzyk, J. W. Freeland, J. Chakhalian, *J. Phys. D: Appl. Phys.* **46,** 385303 (2013); S. D. Han, M. Otaki, R. Jaramillo, A. Podpirka, S. Ramanathan, *J. Solid State Chem*. **190**, 233 (2012).

[21]F. Y. Bruno, K. Z. Rushchanskii, S. Valencia, Y. Dumont, C. Carrétéro, E. Jacquet, R. Abrudan, S. Blügel, M. Ležaić, M. Bibes, A. Barthélémy, *Phys. Rev. B* **88**, 195108 (2013); R. Jaramillo, S. D. Ha, D. M. Silevitch, S. Ramanathan, *Nat. Phys.* **10**, 304 (2014).

[22]S. D. Ha, B. Viswanath, S. Ramanathan, *J. App. Phys.* **111**, 124501 (2012); N. Shukla, T. Joshi, S. Dasgupta, P. Borisov, D. Lederman, S. Datta, *Appl. Phys. Lett.* **105**, 012108 (2014).

[23]See supplemental material at [URL will be inserted by AIP] for a detailed description of the structural characterization of the samples, of the experimental methods and calculations.

[24]M. Imada, A. Fujimori, Y. Tokura, *Rev. Mod. Phys.* **70**, 1039 (1998).

[25]F. Conchon, A. Boulle, R. Guinebretière, C. Girardot, S. Pignard, J. Kreisel, F. Weiss, E. Dooryhée, J-L. Hodeau, *App. Phys. Lett*. **91**, 192110 (2007); F. Conchon, A. Boulle, R. Guinebretiere, E. Dooryhee, J-L. Hodeau, C. Girardot, S. Pignard, J. Kreisel**,** F. Weiss, J. *Phys.: Condens. Matter* **20** ,145216, (2008); F. Conchon, A. Boulle, R. Guinebretière, E. Dooryhée, J.-L. Hodeau, C. Girardot, S. Pignard, J. Kreisel, F. Weiss, L. Libralesso, T. L. Lee, *J. App. Phys.* **103**, 123501 (2008).

[26]G. H. Aydogdu, S. D. Ha, B. Viswanath, S. Ramanathan, *J. App. Phys.* **109**, 124110 (2011).

[27]D. G. Hawthorn, F. He, L. Venema, H. Davis, A. J. Achkar, J. Zhang, R. Sutarto, H. Wadati, A. Radi, T. Wilson, G. Wright, K. M. Shen, J. Geck, H. Zhang, V. Novák, G. A. Sawatzky, *Rev. of Sci. Instr.* **82**, 073104 (2011).

[28]J. P. Hill, D. F. McMorrow, *Acta Crystallogr., Sect. A: Found. Crystallogr.* **52**, 236 (1996).

[29]V. Scagnoli, U. Staub, Y. Bodenthin, M. García-Fernández, A. M. Mulders, G. I. Meijer, G. Hammerl, *Phys. Rev. B* **77**, 115138 (2008).

[30]V. Scagnoli, U. Staub, A. M. Mulders, M. Janousch, G. I. Meijer, G. Hammerl, J. M. Tonnerre, N. Stojic, *Phys. Rev. B* **73**, 100409(R) (2006).